\documentclass{JHEP3}
\usepackage{graphicx}

\def\Tr{\mbox{Tr}\,}

\def\hbar{\hspace{0pt}\raisebox{1pt}{$-$} \hspace{-7pt} h}

\def\5{\overline 5}

\newcommand{\be}{\begin{equation}}
\newcommand{\ee}{\end{equation}}
\newcommand{\bea}{\begin{eqnarray}}
\newcommand{\eea}{\end{eqnarray}}
\newcommand{\nn}{\nonumber}

\title{Causal vs. Analytic constraints on anomalous quartic gauge couplings}
\date{\today
}
\author{L.~Vecchi\\
INFN, Sezione di Trieste and\\
Scuola Internazionale Superiore di Studi Avanzati\\
via Beirut 4, I-34014 Trieste, Italy}
\abstract{
We derive one loop constraints on the anomalous quartic gauge
couplings using a general non-forward dispersion relation for the
elastic scattering amplitude of two longitudinally polarized
vector bosons. We show that for exactly chiral theories more stringent bounds can be obtained by
the assumption that the underlying theory satisfies the causality
principle of Special Relativity.
}
\maketitle
\begin{document}
\section{Introduction}

The general structure of an effective lagrangian is dictated by
the interplay between quantum mechanics, Poincar{\'e} invariance,
and internal symmetries. Its coefficients are not constrained by
the symmetries and must be determined by experiments. Unitarity
usually sets an upper bound on the energy scale below which a
perturbative effective approach is reliable.

We can interpret the standard model (SM) as an effective theory
extending its lagrangian to include new non-renormalizable
operators with unknown coefficients. Some of them enter the
scattering amplitudes of longitudinally polarized vector bosons.
These are called anomalous quartic gauge couplings since they
measure the deviation from the SM predictions. These coefficients
are necessarily connected with the not yet observed Higgs sector.
In the case the Higgs boson is not a fundamental state, or even no
Higgs boson will be observed, they provide important informations
on the nature of the electro-weak symmetry breaking sector.
Whereas there are no significant experimental bounds on them at
the moment~\cite{eboli}, theoretical arguments can reduce
significantly their allowed range and can serve as a guide for
future experiments.

The authors of~\cite{rattazzi} have noticed that the coefficients
of a general effective lagrangian may be constrained by requiring
the S-matrix of the full theory respects some desirable property
such as analyticity, crossing symmetry, Lorentz invariance and
unitarity.

We follow these authors and consider the SM $SU(2)\times
U(1)\rightarrow U(1)$ breaking pattern in the case there exists a
light Higgs-like boson as well as in the case no Higgs boson can
propagate under the cut off of the effective theory. We show that
a general non-forward dispersion relation leads to a less
constraining bound than the one derived by the request the UV
completion respects the causality principle of Special Relativity.
This is not surprising because it is commonly believed that the
analytical properties of the S-matrix are a consequence of its
causal nature.

\section{Analytical bounds}

We briefly review an analytic tool which has been used in the
context of the chiral lagrangian of QCD to constrain some
effective coefficients.

Consider a multiplet of scalar particles, which to be definite we
call pions $\pi^a$, having mass $m$. Assume they are lighter than
any other quanta and that they have appropriate quantum numbers to
forbid the transition $2\pi\rightarrow\pi$. The other states can
be general unstable quanta of complex masses $M$ much greater than
$2m$. The S-matrix element for a general transition
$2\pi\rightarrow2\pi$ is a Lorentz scalar function of the
Mandelstam variables $s,t,u$ and of the mass $m^2$.

We study the amplitude for the elastic scattering
$\pi^a\pi^b\rightarrow\pi^a\pi^b$ and assume it can be
analytically continued to the complex variables $s,t$. We denote
this analytical function by $F(s,t)$ and require that its domain
of analyticity be dictated entirely by the optical theorem and the
crossing symmetry. More precisely, we assume that the
singularities come from simple poles in the correspondence of the
physical masses of the quantum states which can be produced in the
reaction, and branch cuts in the real axis starting at the
threshold of multi-particle production.

Since no mass-less particle exchange is included in $F(s,t)$, the
analytical amplitude satisfies a twice subtracted dispersion
relation for a variety of complex $t$~\cite{martin}. For any non
singular complex point $s,t$ we can write: \bea
\frac{1}{2}\frac{d^2F(s,t)}{ds^2}+P =
\int_{4m^2}^\infty\frac{dx}{\pi}\left\{\frac{ImF(x+i\varepsilon,t)}{(x-s)^3}
+ \frac{ImF_u(x+i\varepsilon,t)}{(x-u)^3}\right\} \label{disp}
\eea where we defined $u=4m^2-s-t$ and used the crossing symmetry
to write the amplitude in the u-channel  as
$F_u(x,t)=F(4m^2-x-t,t)$.

The $P$ on the left hand side of~(\ref{disp}) denotes the second
derivative of the residues. By the analyticity assumption this
term comes entirely from the complex simple poles produced by the
exchange of unstable states. In our discussion the pole term can
be neglected since its contribution turns out not to be relevant .

In the case of forward scattering ($t=0$) the imaginary part
$ImF(x,0)$ is proportional to the total cross section of the
transition $2\pi\rightarrow$'everything' and is therefore non
negative. The crossing symmetry leads to a similar result for the
u-channel. We conclude that $F''(s,0)$ is a strictly positive
function for any real center of mass energy $s$ in the range
$0\leq s\leq4m^2$.

The analyticity assumption can be used to generalize the domain of
positivity of the imaginary part of the amplitude. This can be
seen by expanding $ImF(x+i\varepsilon,t)$ in partial waves in the
physical region and observing that, due to the optical theorem and
the properties of the Legendre polynomials, any derivative with
respect to $t$ at the point $x\geq4m^2$, $t=0$ is non negative.
The Taylor series of $ImF(x+i\varepsilon,t)$ for $t\geq0$ is
therefore greater than zero. Since an analog result holds for the
u-channel, we conclude that the second derivative $F''(s,t)$ is
strictly positive (and analytical) for any real kinematical invariant belonging to the triangle $\Delta=\left\lbrace s,t,u|\,0\leq s,t,u\leq 4m^2\right\rbrace$.

In QCD, the scattering of pions at a scale comparable with their
masses is very well described by the chiral lagrangian. The 4 pion
operators produce order $s^2$ corrections to the scattering
amplitude and eq.~(\ref{disp}) implies positive bounds on some
combination of their coefficients (see~\cite{truong}, for
example).

\subsection{Application to the gauged chiral lagrangian}

We can think of the SM as an effective theory and extend its
action to include non renormalizable operators in the standard
way~\cite{CWZ}.

The anomalous quartic gauge couplings enter the scattering
amplitude of two longitudinally polarized gauge bosons at order
$s^2$. We expect that the method outlined in the previous section
may be used to bound these coefficients.

There exists, however, a fundamental difference from the QCD case.
The assumptions made to derive the relation~(\ref{disp}) are the
analytic, Lorentz and crossing symmetric nature together with the
asymptotic behavior of the amplitude $F(s,t)$. A sufficient
condition for the latter hypothesis to hold is that no massless
particle exchange contribute to $F$ (Froissart bound). In the
electroweak case this latter assumption is not natural because of
the presence of the electromagnetic interactions.

Although we may consider only amplitudes with no single photon
exchange (like $W^\pm Z^0\rightarrow W^\pm Z^0$ for example),
there is still an operative difficulty due to the fact that the
amplitude $F$ is generally dominated by the SM graphs at low
energy scales. These latter give rise to positive contributions to
$F(s,t)$, since the SM is well defined even for vanishing
coefficients, and one is lead to conclude that eq.~(\ref{disp})
implies that the effective operators involved cannot produce a
"too large and negative" contribution to the amplitude $F(s,t)$
and that, as a consequence, no significant bound can be derived in
the gauged theory. Notice that this is also true in the absence of
a light Higgs boson as far as the CM energy is of the order of the
$Z^0$ mass.

One way to overcome these apparent complications is considering
amplitudes with no single photon exchange and evaluating them at a
high scale $s\gg m_Z^2$ with the equivalence theorem (ET). In this
case one has to prove the positivity of the second derivative of
the amplitude is guaranteed in the energy regime in which the approach is defined~\cite{distler}.

Another way, which we decide to follow, is working in the global
limit. The crucial observation in order to justify this assumption
is that in the matching between the effective lagrangian and the
UV theory the transverse gauge bosons contribute, because of their
weak coupling, in a subdominant way to the effective coefficients
of our interest. An accurate estimate of them, and the respective
bounds, can therefore be obtained neglecting completely the gauge
structure and studying the coefficients of the global theory.

Using this conceptually different (though operationally
equivalent) perspective we can study any two by two elastic
scattering amplitude and generalize the analysis of~\cite{distler}
to non-forward scattering.

\subsection{Derivation of the analytical bounds}

We first specialize to the case there appears no Higgs-like boson
under a cut off $\Lambda$.

In this context the basic tool is a non linearly realized
effective lagrangian for the breaking pattern $SU(2)\times
U(1)\rightarrow U(1)$ written in terms of a $SU(2)$ matrix $U =
\exp(i\pi^a \sigma^a/v)$, where $\sigma^a$ are the three Pauli
matrices with $a=1,2,3$ and $v\simeq250$ GeV is the EW vacuum. As
usual, under a global $SU(2)_L\times U(1)_Y$ transformation
$U\rightarrow LUR^\dagger$, where $L \in SU(2)_L$ and $R \in
U(1)_Y \subset SU(2)_R$.

Assuming $m_Z^2\ll\Lambda^2$ and working at energies comparable
with the $Z^0$ mass, the most general lagrangian respecting the
above symmetries and up to $O(s^2)$ is given in
reference~\cite{appelquist}. The globally symmetric version is:
\bea
 {\cal L}_{\rm EWChL} &=&
 -\frac{v^2}{4} \Tr (V_{\mu}V^{\mu}) + \frac{1}{4}\beta_1g^2v^2[Tr(TV_\mu)]^2 \nn\\
               &+& \alpha_{4} [ Tr(V_{\mu}V_{\nu})] ^2 + \alpha_{5} [ Tr(V_{\mu}V^{\mu})] ^2 + \alpha_6Tr(V_\mu V_\nu)Tr(TV^\mu)Tr(TV^\nu) \nn\\
               &+& \alpha_7Tr(V_\mu V^\mu)Tr(TV_\nu)Tr(TV^\nu) + \frac{1}{2}\alpha_{10}[Tr(T V_\mu)Tr(TV_\nu)]^2,
               \label{lag}
                \eea
where $V_\mu = (\partial_{\mu}U)U^\dagger$ and $T = U\sigma^{3}U^\dagger$.

We stress that in this idealized scenario the $\pi^a$ are exact
Goldstone bosons. To avoid any complication with the asymptotic
behavior of the amplitude we can introduce by hand a $\pi^a$ mass
and proceed as in QCD. This mass is actually the consequence of an
explicit symmetry breaking term in the UV theory. Being interested
in constraining the underlying symmetric theory we are forced to
take $m^2\ll m_Z^2,s$. The bounds we derive differ from the QCD ones for this very reason. 

Although no mass gap is present in this context, an approximate
positive constraint for $F''(s,t)$ can be derived. This we do by
noticing that a general dispersion relation like~(\ref{disp}) can
be used to bound the anomalous quartic couplings only if the
$O(s^3)$ contribution to $F(s,t)$ is negligible. In this regime
the second derivative $F''(s,t)$ is dominantly $s$ independent
and, for a small non vanishing imaginary part for $s$, the
dispersion relation can be approximated as: \bea
\frac{1}{2}\frac{d^2F(s,t)}{ds^2} \simeq
\int_0^\infty\frac{dx}{\pi}\left\{\frac{ImF(x+i\varepsilon,t)}{x^3}
+ \frac{ImF_u(x+i\varepsilon,t)}{x^3}\right\} \left(1+
O\left(\frac{s,t}{\Lambda^2} \right)\right) \label{approx} \eea
where the limit $m^2/s\rightarrow0$ was assumed and the resonant
pole term has been neglected. Eq.~(\ref{approx}) shows that, as
far as $O(s^3)$ are negligible compared to $O(s^2)$, the second
derivative of the amplitude is strictly positive.

Before evaluating the bounds we notice that the smallness of the
EW precision tests T parameter~\cite{peskin} is conveniently
achieved by assuming the existence of an approximate global
$SU(2)_C$ custodial symmetry under which the Goldstone boson
matrix transforms as the adjoint representation. The dominant
coefficients associated to anomalous quartic gauge operators are
$\alpha_4$ and $\alpha_5$ and any $\pi^a\pi^b \rightarrow
\pi^c\pi^d$ scattering amplitude can be written in terms of a
function $A(s,t,u)$. The relevant processes turn out to be: \bea
A(\pi^0\pi^0 \longrightarrow \pi^0\pi^0) &=& A(s,t,u) + A(t,s,u) + A(u,t,s) \nn \\
A(\pi^\pm\pi^0 \longrightarrow \pi^\pm\pi^0) &=& A(t,s,u),
\label{physA} \eea where, at one loop level and in the limit
$m^2/s\rightarrow0$, we have~\cite{GL} \bea A(s,t,u) &=&
\frac{s}{v^2} + \frac{4}{v^4} \left[
2\alpha_{5}(\mu)s^2+\alpha_{4}(\mu)(t^2+u^2) +
\frac{1}{(4\pi)^2}\frac{10s^2 + 13(t^2+u^2)}{72}\right] \nn \\
&-& \frac{1}{96\pi^2v^4} \left[ t(t-u)\log\left(
-\frac{t}{\mu^2}\right)  + u(u-t)\log\left(
-\frac{u}{\mu^2}\right)  + 3s^2\log\left( -\frac{s}{\mu^2}\right)
\right]. \eea Notice that we have chosen to work with the
renormalized coefficients $\alpha_{4,5}(\mu)$ as defined by the
modified minimal subtraction scheme, rather than using the non
standard normalization of~\cite{GL}.

We can now derive~(\ref{physA}) twice with respect to $s$ and
evaluate the result at $s+i\varepsilon,t$, where $0<s,t\ll\Lambda^2$. It is
convenient to choose a different representation for the
kinematical invariants in order to eliminate the logarithms in the
final result. We define a scale $w=\sqrt{s(s+t)}=\sqrt{-su}>s$ and
obtain: \bea
\alpha_{4}(w) + \alpha_{5}(w) &>& -\frac{1}{16}\frac{1}{(4\pi)^2}\nn\\
\alpha_{4}(w)  &>&
\frac{1}{12}\frac{1}{(4\pi)^2}\left(-\frac{7}{6}+\frac{1}{8}\left(\frac{w}{s}+\frac{s}{w}\right)^2\right).
\label{anaM} \eea For $t=0$ we have
$\alpha_4+\alpha_5\gtrsim-0.40\times10^{-3}$ and
$\alpha_4\gtrsim-0.35\times 10^{-3}$ at an arbitrary scale
$w=s\ll\Lambda^2$. This result coincides with the one obtained
in~\cite{distler}, as expected.

In the case of non-forward scattering, the bound on $\alpha_4(w)$
cannot get arbitrarily large (large $w$ or,
equivalently, large $t$) because at some unknown scale, much
smaller than $\Lambda^2$, the $O(s^3)$ corrections become relevant
in the determination of the amplitude and the bound would not
apply. Without a detailed knowledge of the perturbative expansion
in the weak coupling $s/\Lambda^2$, (that is, of the full theory!)
we cannot realistically tell which is the strongest bound derived
by this analysis.

What we can certainly do is to compare~(\ref{anaM}) with the well known constraints on the corresponding parameters $l_1=4\alpha_5$ and $l_2=4\alpha_4$ of QCD. Strong bounds on these coefficients have been evaluated in the triangle $\Delta$~\cite{Wanders}. We may interpret our analysis as a study of the axiomatic constraints on the two pion amplitudes in the complementary region $m^2\ll s\ll\Lambda^2$. Using the notation introduced in~\cite{GL} we translate~(\ref{anaM}) into $2\bar l_1+4\bar l_4\gtrsim3$ and $\bar l_2\gtrsim0.3$. These constraints are compatible with the experimental observations~\cite{Bij} but are less stringent than those obtained in ~\cite{Wanders}. 

We conclude that our analysis does not lead to an improvement of the bounds on $\bar l_{1,2}$.
If the chiral symmetry is exact, on the other hand, eqs.~(\ref{anaM}) represent stringent bounds on the anomalous quartic couplings implied by the assumptions of analyticity, crossing symmetry, unitarity and Lorentz invariance of the S-matrix.

Eq.~(\ref{approx}) is not rigorous if a light state
enters the processes under consideration and therefore~(\ref{anaM}) are not valid if a Higgs-like scalar propagates under the cutoff. In the next paragraph we
discuss an approach which works in this context as well, provided the chiral symmetry is exact.

\section{Causal bounds}

Given a general solution of the equations of motion derived
from~(\ref{lag}) we can study the oscillations around it.
Consistency with Special Relativity requires the oscillations to
propagate sub-luminally. This request may be expressed as a
constraint on the same coefficients which enter the elastic
scattering of two Goldstone bosons because the dynamics of the
oscillation on the background can be interpreted as a scattering
process on a macroscopic `object`. If the background has a
constant gradient, the presence of super-luminal propagations sum
up and can in principle become manifest in the low energy
regime~\cite{rattazzi}.

A constant gradient solutions admitted by the
lagrangian~(\ref{lag}) is defined by $\pi_{0}=\sigma C_\mu x^\mu$,
where $\sigma$ is a generic isospin direction and the constant
vector $C_\mu$ is fine-tuned in order to satisfy $C^2\ll v^4$. The
quadratic lagrangian for the oscillations $\delta\pi=\pi-\pi_0$
around the background have the general form: \bea {\cal
L}=\delta\pi\left(p^2+ \frac{\alpha}{v^4}\left(
Cp\right)^2\right)\delta\pi, \label{EOM} \eea with
$\alpha=\alpha_{4},\alpha_{4}+\alpha_{5}$. In the evaluation
of~(\ref{EOM}) we neglected $O(Cx/v)$ terms. We can imagine in
fact the non trivial background to be switched on in a finite
space-time domain so that the latter approximation is seen as a
consequence of the fine-tuning of the parameter $C_\mu$.

A perturbative study of the interacting field $\delta\pi$ is in
principle possible for energies under a certain scale (to be
definite we call this scale the cut-off of the effective theory).
By assumption, this cut off is arbitrarily close to $\Lambda$ as
$C^2/v^4$ goes to zero and, having this fact in mind, we simply
denote it as $\Lambda$.

A necessary condition for such a perturbative study to make any
sense is that the quadratic lagrangian be well defined. This is
the case for~(\ref{EOM}) only if $\alpha\geq0$. In fact, the field
$\delta\pi$ has velocity $dE/dp=E/p$ (where $p^\mu=(E,\bar p)$ and
$|\bar p|=p$) and for $\alpha<0$ its quanta propagate
super-luminally.

It is important to notice that the presence of super-luminal modes
is not the consequence of a bad choice of the vacuum. The
quadratic hamiltonian is stable in any vacuum (parametrized by
$C_\mu$) if $\alpha$ is 'sufficiently small' but generally leads
to violations of the causality principle of Special Relativity
when $\alpha<0$. In the latter hypothesis then different inertial
frames may not agree on the physical observations and, for
example, the quadratic hamiltonian may appear unbounded from below
to a general Lorentzian frame boosted with a sufficiently high
velocity.

We finally interpret the constraint $\alpha\geq0$ as a causal bound.

The effective coefficients $\alpha$ which appear in the
perturbative analysis are actually the renormalized couplings so
that the above bound can be extended to all energy scales
$w<\Lambda^2$, where the perturbative study is assumed to be
meaningful, after taking into account the running effect: \bea
\alpha_{4}(w) + \alpha_{5}(w) \geq  \frac{1}{8}\frac{1}{(4\pi)^2}\log\left( \frac{\Lambda^2}{w}\right)  \nn \\
\alpha_{4}(w) \geq \frac{1}{12}\frac{1}{(4\pi)^2}\log\left(
\frac{\Lambda^2}{w}\right). \label{cau} \eea

This approach may be applied even to scenarios in which a scalar
Higgs, composite or fundamental, can propagate under the cut off.
In this latter case the causal constraints read $\alpha_4\geq0$
and $\alpha_4+\alpha_5\geq0$ but now the coefficients do not have
any scale dependence because the theory has no extra-SM
divergences at order $s^2$. Therefore, the possibility
$\alpha_4=\alpha_5=0$ can not and must not be excluded (consider
the particular example of the SM). The analytical bounds, which
would imply a strict inequality, do not apply as already noticed.

The bounds~\ref{cau} cannot be compared to the QCD ones because $\pi_{0}$ does not solve the equations of motion when $m\neq0$.

\section{Conclusions}

We have derived general bounds on the anomalous quartic gauge
couplings using two distinct approaches. The causal one relies on
the absence of superluminal propagations. The analytical one relies on the assumption of
analyticity, crossing and Lorentz symmetry together with a good
behavior at infinity of the scattering amplitude $F(s,t)$. The
latter method works in the context of a strongly coupled theory with
no Higgs propagating at low energy only. In this scenario~(\ref{anaM})
can be compared to~(\ref{cau}). We see that the bound on
$\alpha_4+\alpha_5$ is clearly dominated by the causal result and
that this is also the case for $\alpha_4$ if, roughly, the ratio
$(w/s)^2$ does not exceed $16\log(\Lambda/\sqrt{w})$. We cannot
tell if the analytical bound still apply up to this scale

More importantly, if the fermionic effects are considered
separately from $\alpha_{4,5}$, a realistic estimate of the
constraints should take the fermions couplings to the Goldstone
bosons into account. It is easy to see that the one loop effect
induced by the SM fermions gives rise to a positive contribution
to the second derivative of the amplitude. This of course lowers
the analytical bounds while the causal argument remains valid
and~(\ref{cau}) is not altered.

The bound~(\ref{cau}) for the higgsless
scenario, together with the constraint $\alpha_4\geq0$ and
$\alpha_4+\alpha_5\geq0$ for the light Higgs-like scenario provide
the most stringent and reliable bounds on the effective coefficients
$\alpha_{4,5}$.

In order to have a rough estimate of~(\ref{cau}) we assume
$\Lambda\sim1$ TeV and get $\alpha_4+\alpha_5\gtrsim 3.8\times
10^{-3}$, $\alpha_4\gtrsim 2.5\times 10^{-3}$ at the $Z^0$ pole.
These values lie inside the very wide experimental bounds
$-0.1\lesssim\alpha_{4,5}\lesssim0.1$. Eqs.~(\ref{cau}) significantly reduce the
allowed range. 

The experimental constraints are extremely weak
since they have been derived by estimating the loop corrections
induced by $\alpha_{4,5}$ on the electroweak precision
parameters~\cite{eboli}. A direct
measurement of the anomalous gauge couplings turns out to be of fundamental importance in order to have some insight on the actual nature of the electroweak breaking sector~\cite{FV}.
LHC may improve the bounds~\cite{eboli} by an order of magnitude but the linear
collider seems far more appropriate to resolve the
coefficients~\cite{LC}. The measurement of a negative value of
$\alpha_4$ and $\alpha_4+\alpha_5$ at the next linear collider
would therefore signal a breaking of causality, irrespective of
the presence of a light scalar state like the Higgs boson. This
seems a rather unlikely possibility because it would require too
drastic a modification of our physical understanding. A more
conservative point of view consists in interpreting the
bounds~(\ref{cau}) as theoretical constraints on the full theory.

\acknowledgments
This work is partially supported by MIUR and the RTN European Program MRTN-CT-2004-503369.



\begin{thebibliography}{99}


\bibitem{eboli}
O.~J.~P.~Eboli, M.~C.~Gonzalez-Garcia and J.~K.~Mizukoshi,
  Phys.\ Rev.\  D {\bf 74}, 073005 (2006)
  [arXiv:hep-ph/0606118].


  H.~J.~He, Y.~P.~Kuang and C.~P.~Yuan,
  Phys.\ Rev.\  D {\bf 55}, 3038 (1997)
  [arXiv:hep-ph/9611316];\\
  A.~S.~Belyaev, O.~J.~P.~Eboli, M.~C.~Gonzalez-Garcia, J.~K.~Mizukoshi, S.~F.~Novaes and I.~Zacharov,
  Phys.\ Rev.\  D {\bf 59}, 015022 (1999)
  [arXiv:hep-ph/9805229].

\bibitem{rattazzi}
  A.~Adams, N.~Arkani-Hamed, S.~Dubovsky, N.~Nicolis, R.~Rattazzi,
  JHEP \ 0610:014,(2006)

\bibitem{martin}
  A.~Martin,
  Nuovo\ Cim.\ A {\bf 42}, 930 (1966)

\bibitem{truong}
T.~N.~Pham and T.~N.~Truong,
  Phys.\ Rev.\  D {\bf 31}, 3027 (1985).

\bibitem{CWZ}
C.~G.~.~Callan, S.~R.~Coleman, J.~Wess and B.~Zumino,
  Phys.\ Rev.\  {\bf 177}, 2247 (1969).


\bibitem{distler}
J.~Distler, B.~Grinstein, R.~A.~Porto and I.~Z.~Rothstein,
  Phys.\ Rev.\ Lett.\  {\bf 98}, 041601 (2007)
  [arXiv:hep-ph/0604255].


\bibitem{appelquist}
T.~Appelquist and G.~H.~Wu,
  Phys.\ Rev.\  D {\bf 48}, 3235 (1993)
  [arXiv:hep-ph/9304240].

\bibitem{peskin}
M.~E.~Peskin and T.~Takeuchi,
  Phys.\ Rev.\  D {\bf 46}, 381 (1992).


\bibitem{GL}
J.~Gasser and H.~Leutwyler,
  Annals Phys.\  {\bf 158}, 142 (1984).


\bibitem{Wanders}
  B.~Ananthanarayan, D.~Toublan and G.~Wanders,
  Phys.\ Rev.\  D {\bf 51}, 1093 (1995)
  [arXiv:hep-ph/9410302].


\bibitem{Bij}
  J.~Bijnens,
  Prog.\ Part.\ Nucl.\ Phys.\  {\bf 58}, 521 (2007)
  [arXiv:hep-ph/0604043].





\bibitem{LC}
E.~Boos, H.~J.~He, W.~Kilian, A.~Pukhov, C.~P.~Yuan and
P.~M.~Zerwas,
  Phys.\ Rev.\  D {\bf 61}, 077901 (2000)
  [arXiv:hep-ph/9908409].


\bibitem{FV}
  M.~Fabbrichesi and L.~Vecchi,
  Phys.\ Rev.\  D {\bf 76}, 056002 (2007)
  [arXiv:hep-ph/0703236].












 \end{thebibliography}
 \end{document}